\documentclass[article,nojss,shortnames]{jss}

\usepackage{orcidlink,thumbpdf,lmodern}
\usepackage{amsmath}
\usepackage{amssymb}
\usepackage{mathtools}
\usepackage{commath}

\usepackage{framed}

\newcommand{\fct}[1]{\code{#1()}}

\author{Leonardo N. Ferreira~\orcidlink{0000-0003-1445-0590}\\Max Planck Institute for Human Development
}
\Plainauthor{Leonardo N. Ferreira}

\title{From Time Series to Networks in \proglang{R} with\\the \pkg{ts2net} Package}
\Plaintitle{From Time Series to Networks in R with ts2net}
\Shorttitle{From Time Series to Networks in \proglang{R}}

\Abstract{
  Network science established itself as a prominent tool for modeling time series and complex systems. This modeling process consists of transforming a set or a single time series into a network. Nodes may represent complete time series, segments, or single values, while links define associations or similarities between the represented parts. \proglang{R} is one of the main programming languages used in data science, statistics, and machine learning, with many packages available. However, no single package provides the necessary methods to transform time series into networks. This paper presents \pkg{ts2net}, an \proglang{R} package for modeling one or multiple time series into networks. The package provides the time series distance functions that can be easily computed in parallel and in supercomputers to process larger data sets and methods to transform distance matrices into networks. \pkg{Ts2net} also provides methods to transform a single time series into a network, such as recurrence networks, visibility graphs, and transition networks. Together with other packages, \pkg{ts2net} permits using network science and graph mining tools to extract information from time series.
}

\Plainkeywords{time series, networks, graphs, network science, complex networks, R}

\Address{
  Leonardo N. Ferreira\\
  Center for Humans \& Machines\\
  Max Planck Institute for Human Development\\
  Lentzeallee 94, 14195 Berlin, Germany\\
  E-mail: \email{ferreira@leonardonascimento.com}\\
  URL: \url{https://www.leonardoferreira.com/}
}

\begin{document}



\section[Introduction]{Introduction} 
\label{sec:introduction}

Networks are one of the most prominent tools for modeling complex systems \citep{mitchell06}. Complex systems are commonly represented by a set of time series with interdependencies or similarities \citep{silva18}. This set can be modeled as a network where nodes represent time series, and links connect pairs of associated ones. Associations are usually measured by time series distance functions, carefully chosen by the user to capture the desired similarities. Thus, the network construction process consists of two steps. First, the distance between all pairs of time series is calculated and stored in a distance matrix. Then, this distance matrix is transformed into an adjacency matrix. The network topology permits not just the analysis of the small parts (nodes) in the system but also their relationship (edges). This powerful model allows network science and graph mining tools to extract information from different domains such as climate science \citep{boers19,ferreira21a}, social science \citep{ferreira21b}, neuroscience \citep{bullmore09}, and finance \citep{tumminello10}.

In recent years, networks have also been applied to model a single time series. The goal is to transform one time series into a network and apply network science techniques to perform time series analysis \citep{silva21}. Different modeling approaches have been proposed, and the general idea is to map single values \citep{lacasa08}, time windows, transitions \citet{campanharo11}, or recurrence \citet{donner10} as nodes and their relationship by links. Then, the general goal is to use network science and graph mining tools to extract information from the modeled time series.

\proglang{R} is one of the most used programming languages in statistics, data science, machine learning, complex networks, and complex systems. Different packages provide distance functions and network analysis, but no single package provides the necessary tools to model one or multiple time series as networks. This paper presents \pkg{ts2net}, a package to construct networks from time series in \textsf{R}.\ \pkg{Ts2net} permits the use of network science to study time series. Together with other \proglang{R} packages, \pkg{ts2net} permits distance calculations, network construction, visualization, statistical measures extraction, and pattern recognition using graph mining tools. Most of the key functions are implemented using multi-core programming, and some useful methods make \pkg{ts2net} easy to run in supercomputers and computer clusters. The package is open-access and released under the MIT license.

In the following sections, I present some background concepts related to time series network modeling (Sec.~\ref{sec:background}), the \pkg{ts2net} package (Sec.~\ref{sec:ts2net}), and two applications using real data (Sec.~\ref{sec:applications}), followed by some final remarks. 






\section[Background Concepts]{Background Concepts} 
\label{sec:background}

In summary, the whole network construction process involves the calculations of time series distance functions, represented by a distance matrix, and the transformation of this matrix into a network. This section presents the distances and the network construction methods available in the \pkg{ts2net} package. 

\subsection{Distance functions}
\label{seubsec:dist_funcs}

Time series comparison is a fundamental task in data analysis. This comparison can estimated using similarity or distance functions \citep{deza09,esling12,wang13}. Some research domains use the concept of similarity, while others prefer distance functions. Similarity and distance are inverse concepts that have the same usefulness. Here in this paper, for simplicity, I use the distance concept. However, it is important to remember that a similarity function can be generally transformed into distance functions and vice-versa. The distance or metric $d : \mathcal{X} \times \mathcal{X} \to \mathbb{R}$ is a function that for all time series $X,Y,Z \in \mathcal{X}$, the following properties hold:

\begin{itemize}
    \item \textit{non-negativity}: $d(X,Y) \geq 0 $, 
    \item \textit{symmetry}: $d(X,Y) = d(Y, X)$, 
    \item \textit{reflexity}: $d(X,Y) = 0 \iff X=Y$,
    \item \textit{triangular difference}: $d(X,Y) \leq d(X,Z) + d(Z,Y)$. 
\end{itemize}

Some of the time series comparison measures used in \pkg{ts2net} are not true metrics  because they do not follow one or more required properties above \citep{deza09}. However, for simplicity reasons, this paper uses the term ``distance'' to generally refer to any of these functions. The rest of Sec.~\ref{seubsec:dist_funcs} reviews the main distance functions implemented in the \pkg{ts2net} package.

\subsubsection{Correlation-based distances}

The Pearson Correlation Coefficient (PCC) $r$ is one of the most used association measures \citep{asuero06}. Given two time series $X$ and $Y$ (both with length $T$), and their respective mean values $\overline{X}$ and $\overline{Y}$, the PCC is defined as

\begin{equation}
    \label{eq_pcc}
    r(X,Y) = \frac{{}\sum_{i=1}^{T} (X_i - \overline{X})(Y_i - \overline{Y})}{\sqrt{\sum_{i=1}^{T} (X_i - \overline{X})^2(Y_i - \overline{Y})^2}}.
\end{equation}

The PCC can be transformed into a distance function by considering the absolute value of $r$ as follows:

\begin{equation}
    \label{eq:d_cor_abs}
    d_{\text{cor\_abs}}(X,Y) = 1 - |r(X,Y)|.
\end{equation}

\noindent In this case (Eq.~\ref{eq:d_cor_abs}), both strong correlations ($r \approx 1$) and strong anti-correlation ($r \approx -1$) lead to $d_{cor\_abs}(X,Y) \approx 0$. This distance is particularly useful if both correlated and anti-correlated time series should be linked in the network. Conversely, if only strong positive correlations or strong negative correlation should be mapped into links in the network, then Eq.~\ref{eq:d_cor_pos} or Eq.~\ref{eq:d_cor_neg} can be used instead. 
\begin{equation}
    \label{eq:d_cor_pos}
    d_{\text{cor\_pos}}(X,Y) = 1 - \max(0,\ r(X,Y)) 
\end{equation}
\begin{equation}
    \label{eq:d_cor_neg}
    d_{\text{cor\_neg}}(X,Y) = 1 - \max(0,\ -r(X,Y)) 
\end{equation}
Cross-correlation is also a widely used measure to find lagged associations between time series. Instead of directly calculating the PCC, the cross-correlation function (CCF) applies different lags (displacement) to the series and then returns the correlation to each lag. Instead of returning a single correlation value, as in the PCC, the CCF returns an array of size $2 \cdot \tau_{max} + 1$, where $\tau_{max} \in \mathbb{N}$ is the maximum lag defined by the user. Each element of this array corresponds to the $r(X, Y_{\tau})$ between the time series $X$ and $Y_{\tau}$, which is the $Y$ lagged by $\tau \in [-\tau_{max},\tau_{max}]$ units.

The CCF can be transformed into a distance function in the same way as the PCC (Eqs.~\ref{eq:d_cor_abs}, \ref{eq:d_cor_pos}, and \ref{eq:d_cor_neg}). The maximum absolute  (Eq.~\ref{eq:d_ccor_abs}), positive (Eq.~\ref{eq:d_ccor_pos}), or negative (Eq.~\ref{eq:d_ccor_neg}) correlations $r(X, Y_{\tau}), \forall \ \tau \in [-\tau_{max},\tau_{max}]$. 

\begin{equation}
    \label{eq:d_ccor_abs}
    d_{\text{ccor\_abs}}(X,Y) = \displaystyle 1 - \bigg(\max_{\tau \in [-\tau_{max},\tau_{max}]}|r(X,Y_{\tau})|\bigg).
\end{equation}

\begin{equation}
    \label{eq:d_ccor_pos}
    d_{\text{ccor\_pos}}(X,Y) = \displaystyle 1 - \bigg(\max_{\tau \in [-\tau_{max},\tau_{max}]} \max(0,\ r(X,Y_{\tau})\bigg)
\end{equation}

\begin{equation}
    \label{eq:d_ccor_neg}
    d_{\text{ccor\_neg}}(X,Y) = \displaystyle 1 - \bigg(\max_{\tau \in [-\tau_{max},\tau_{max}]} \max(0,\ -r(X,Y_{\tau})\bigg)
\end{equation}

One crucial question is the correlation statistical significance. The significance test avoids spurious correlations and false links in the network. One way to test consists of using the Fisher z-transformation defined as \citep{asuero06}:
\begin{equation}
    \label{eq:fisher_transform}
    z(r) = \frac{1}{2}\ln{\frac{1+r}{1-r}} = (\ln(1+r) - \ln(1-r)) / 2.
\end{equation}
This transformation stabilizes the variance and can be used to construct a confidence interval (CI) using standard normal theory, defined (in z scale) as

\begin{equation}
    \label{eq:fisher_confidence_interval_z}
    \Bigg(z_{min} = z - \frac{q}{\sqrt{t-3}}, z_{max} = z + \frac{q}{\sqrt{t-3}}\Bigg),
\end{equation}

\noindent where $q$ is the desired normal quantile and $T$ is the length of the time series. The CI can be transformed back into the original scale using the inverse transformation: 

\begin{equation}
    \label{eq:fisher_inverse}
    r(z) = \frac{e^{2z}-1}{e^{2z}+1},
\end{equation}

leading to 

\begin{equation}
    \label{eq:fisher_confidence_interval_ci}
    \text{CI} = (r(z_{min}), r(z_{max})).
\end{equation}

The correlation (Eqs.~\ref{eq:d_cor_abs}, \ref{eq:d_cor_pos}, and \ref{eq:d_cor_neg}) and cross-correlation distances (Eqs.~\ref{eq:d_ccor_abs}, \ref{eq:d_ccor_pos}, and ~\ref{eq:d_ccor_neg}) can be adapted to consider the statistical test given a significance level $\alpha$. These adaptations consist of substituting the correlation function $r(X,Y)$ in the equations with a binary 

\begin{equation}
    \label{eq:r_test}
    r_{test}(X,Y) = 
        \begin{cases}
          0 & \text{if } r(X,Y) \in \text{CI (} H_0 \text{ not discarded)} \\
          1 & \text{otherwise} \text{ (} H_1 \text{ holds).} 
        \end{cases} 
\end{equation}

\subsubsection{Distances based on information theory}

Mutual information (MI) is one of the most common measures used to detect non-linear relations between time series. The MI measures the average reduction in uncertainty about X that results from learning the value of Y or vice versa \citep{mackay03}. The concept of information is directly linked to Shannon's entropy, which captures the ``amount of information'' in a time series. Given a discrete time series $X$ and its probability mass function $P(X)$, the marginal entropy $H(X)$ is defined as

\begin{equation}
    \label{eq:entropy}
    H(X) = -\sum_{i=1}^{t}P(x_i)\log P(x_i).
\end{equation}

The conditional entropy $H(X|Y)$ measures how much uncertainty $X$ has when $Y$ is known. 

\begin{equation}
    \label{eq:entropy_conditional}
    H(X|Y) = -\sum_{i,j}p(X_i,Y_j)\log \frac{p(X_i,Y_j)}{p(Y_j)}
\end{equation}

A histogram can be used to estimate the probabilities. The number of bins in the histogram directly influences the entropy results and should be chosen accordingly. Some common options to estimate the number of bins are the Sturges's, Scott's, and Freedman-Diaconis criteria \citep{wand97}.

The mutual information (MI) can be estimated using Eq.~\ref{eq:mutual_information}.

\begin{equation}
    \label{eq:mutual_information}
    I(X,Y) = H(X) + H(Y) - H(X,Y)
\end{equation}

The mutual information has only a lower bound (zero), which makes it sometimes difficult to interpret the results. Some normalization methods can limit the MI to the interval $[0,1]$ to make the comparison easier. The normalized mutual information (NMI) can be defined as
\begin{equation}
    \label{eq:norm_mutual_information}
    NMI(X,Y) = I(X,Y)/U,
\end{equation}
where $U$ is the normalization factor. Common normalization approaches are either the half of the sum of the entropies $U=0.5 H(X) H(Y)$, minimum entropy $U=\min(H(X),H(Y))$, maximum entropy $U=\max(H(X),H(Y))$, or the squared root of the entropies product $U=\sqrt{H(X)H(Y)}$ \citep{vinh10}. The NMI distance is

\begin{equation}
    \label{eq:nmi_distance}
    d_{\text{nmi}}(X,Y) = 1 - NMI(X,Y).
\end{equation}

Another common distance function based on MI is the Variation of Information (VoI) (Eq.~\ref{eq:voi}) \citep{vinh10}. VoI, different from the MI, is a true metric.

\begin{equation}
    \label{eq:voi}
    d_{\text{voi}}(X,Y) = H(X) + H(Y) - 2I(X,Y)
\end{equation}

The maximal information coefficient (MIC) is an algorithm that searches for the largest MI between two series \citep{reshef11}. MIC considers that if there exists a relationship between two time series, then it is possible to draw a grid on the scatter plot of these two series that encapsulates that relationship. MIC automatically finds the best grid and returns the largest possible MI.

\subsubsection{Dynamic Time Warping}
\label{subsubsec_dtw}

\begin{figure}[t!]
\centering
\includegraphics[width=1\linewidth]{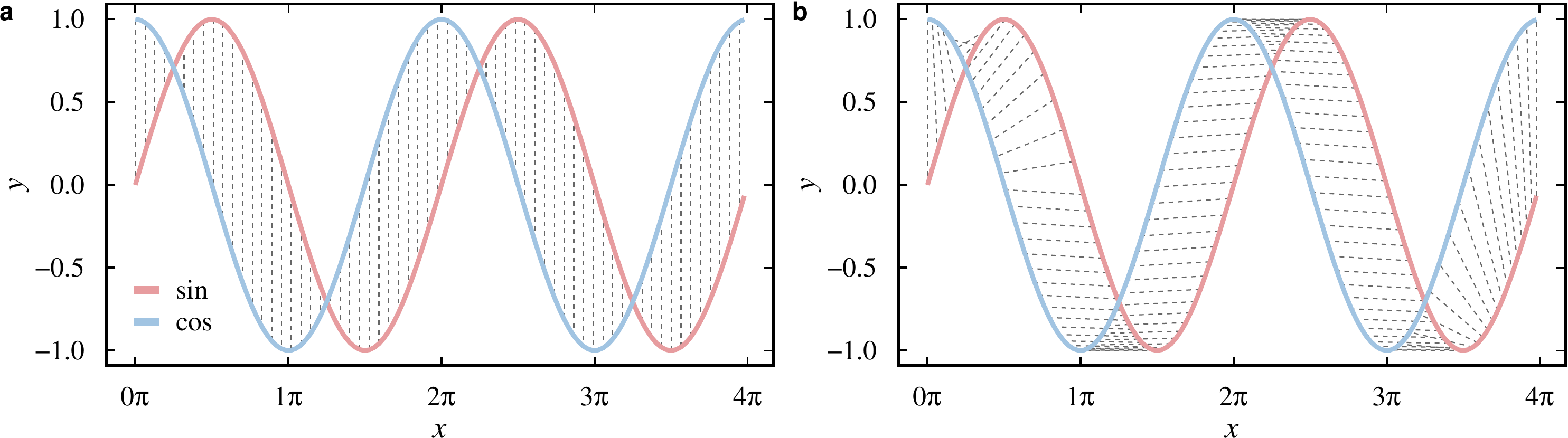}
\caption{\label{fig:dtw} Comparison between lock-step and elastic measures. The figure shows a sine (red) and a cosine (blue) time series. The gray dashed lines show the comparison of the values in (a) a traditional lock-step distance such as the euclidean distance and (b) in an elastic one as in DTW.}
\end{figure}

The Dynamic Time Warping (DTW) \citep{berndt94,giorgino09} is an elastic distance function that uses a warping path to align two time series and then compare them. Different from lock-step measures (such as the euclidean distance), elastic measures use a warping path defining a mapping between the two series (Fig.~\ref{fig:dtw}). The optimal path minimizes the global warping cost. The DTW distance $d_{\text{DTW}}(X,Y) = dtw(i=t,j=t)$ between two time series $X$ and $Y$ (both of length $T$) can be defined using dynamic programming. The total distance is the cumulative distance calculated from the recurrence relation $dtw(i,j)$, defined as:
\begin{equation}
    \label{eq:dtw}
    d_{\text{dtw}}(X,Y) = dtw(i=T,j=T) = 
        \begin{cases}
          \infty & \text{if } i=0 \text{ xor } j=0\\
          0      & \text{if } i=j=0\\
          |X_i-Y_j| + \min
        \begin{cases}
          dtw(i-1,j)  \\
          dtw(i,j-1)  \\
          dtw(i-1,j-1)  \\
        \end{cases}  & \text{otherwise}\\
        \end{cases} 
\end{equation}
\subsubsection{Event-based distances}

\begin{figure}[t!]
\centering
\includegraphics[width=1\linewidth]{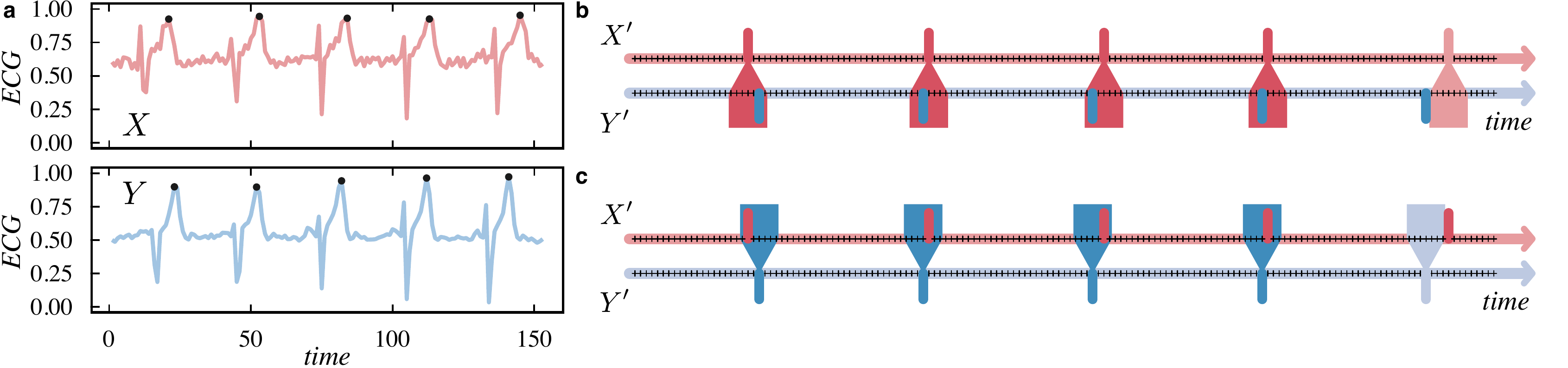}
\caption{\label{fig:events_associations} Intuition behind event-based distance functions. (a) Two ECG time series $X$ and $Y$ are depicted in red (top) and blue (bottom), respectively. Data were extracted and adapted from \citep{physioBank00,liu16}. Black dots represent  marked events in both time series. (b and c) Segments in red ($X^{ev}$) and blue ($Y^{ev}$) represent the events (black dots) in $X$ and $Y$, respectively. (b) For each event in $X^{ev}$, the distance function tries to find another event in $Y^{ev}$ considering a time window ($\tau= \pm 3$), illustrated by the red and blue pentagons. (c) The same search can be performed the other way around.}
\end{figure}

The idea behind event-based distance functions is to compare marked time points where some events occur in the time series. The definition of an event is specific to the domain where the time series comes from. For example, Fig.~\ref{fig:events_associations}-a illustrates two ECG time series $X$ and $Y$ where events (black dots) can be defined as some local maximum values as the T wave representing the ventricular repolarization during the heartbeat. Another example is the concept of extreme events of precipitation that can be defined as the days with rainfall sums above 99th percentiles of wet days during rain season \citep{boers19}. Common approaches used to define events are local maxima and minima, values higher or lower than a threshold, values higher or lower percentiles, or outliers in the series. An event time series $X^{ev}$ is here defined as a binary time series composed of ones and zeros representing the presence or absence of an event respectively in a time series $X$. Fig.~\ref{fig:events_associations}-b and c illustrate these events time series for the ECG time series in Fig.~\ref{fig:events_associations}-a. Another common representation is the events sequence $X'= \{t_1, t_2,\cdots,t_N\}$ that defines the time indices where each event occur.

Event-based distance functions try to find co-occurrences or synchronization between two event time series. The general aim is to count pairs of events in both time series that occur at the same time window of length. For example, considering the five events (black dots) in Fig.~\ref{fig:events_associations}-a, Fig~\ref{fig:events_associations}-b shows the co-occurrence of events from $X^{ev}$ to $Y^{ev}$ considering a time window. For the five events, only the last one does not co-occur. Fig~\ref{fig:events_associations}-c shows the inverse process that can or not be equal depending on the specific distance function. Thus, functions can be symmetrical or not (directed). Symmetrical distances consider co-occurrences from $X$ to $Y$ but also from $Y$ to $X$. Directed measures only consider the events in a precursor time series $X$ compared to another one $Y$, but not the other way around. 

\citet{quiroga02} proposed a method to measure the synchronization between two event time series. Considering two event sequences $X' = \{t_1,\cdots,t_{N^X}\}$ and $Y' = \{t_1,\cdots,t_{N^Y}\}$, and a time lag $\tau$, this method counts the number of times an event appears in $X'$ shortly after it appears in $Y'$ considering using the following equation

\begin{equation}
    \label{eq:quiroga_count1}
    c^{\tau}(X'|Y') = \sum_{i=1}^{N_{X}}\sum_{j=1}^{N_{Y}}J_{ij}^{\tau},
\end{equation}

\noindent where $J_{ij}^{\tau}$ is defined as

\begin{equation}
    \label{eq:quiroga_count2}
    J_{ij}^{\tau} = 
        \begin{cases}
          1   & \text{, if } 0 < t_i^X - t_j^Y \leq \tau \\
          1/2 & \text{, if } t_i^X = t_j^Y \\
          0   & \text{, otherwise}
        \end{cases} 
\end{equation}

\noindent where $t_i^X$ and $t_j^Y$ are the time indices $i$ and $j$ from $X'$ and $Y'$ respectively. Analogously, $c^{\tau}(Y'|X')$ can also be calculated using Eq.~\ref{eq:quiroga_count1}. The symmetrical event-based distance between $X'$ and $Y'$, can be calculated using Eq.~\ref{eq:quiroga_count3}, where $d_{\text{q}}(X',Y') = 0$ indicates full synchronization. 

\begin{equation}
    \label{eq:quiroga_count3}
    d_{\text{es}}^{sy}(X',Y') =  1 - \frac{c^{\tau}(X'|Y') + c^{\tau}(Y'|X')}{\sqrt{N_{X}N_{Y}}}.
\end{equation}

The asymmetrical version counts only the number of events in $X'$ that precede those in $Y'$ (or the opposite in an analogous way), and it is originally defined in the interval [-1,1]. To standardize the interval with the other distances in this paper, Eq.~\ref{eq:quiroga_count4} measures the asymmetrical distance where zero means that all events in $X'$ precede those in $Y'$, and one no synchronization.

\begin{equation}
    \label{eq:quiroga_count4}
    d_{\text{es}}^{as}(X',Y') =  1 - \frac{c^{\tau}(Y'|X') - c^{\tau}(X'|Y') + \sqrt{N_{X}N_{Y}}}{2\sqrt{N_{X}N_{Y}}}
\end{equation}

The authors also propose a local time window $\tau_{ij}$ when event rates changes in the time series. The counting function (Eq.~\ref{eq:quiroga_count1}) can be adapted to a local time window by replacing $\tau$ by 

\begin{equation}
    \label{eq:quiroga_tau_local}
    \tau_{ij} = \min\{ t_{i+1}^X-t_{i}^X, t_{i}^X-t_{i-1}^X, t_{j+1}^Y-t_{j}^Y, t_{j}^Y-t_{j-1}^Y\}/2.
\end{equation}

\citet{boers19} introduce a parameter $\tau_{\text{max}}$ that limits the maximum local time window (Eq.~\ref{eq:quiroga_tau_local}). 

\begin{figure}[t!]
\centering
\includegraphics[width=1\linewidth]{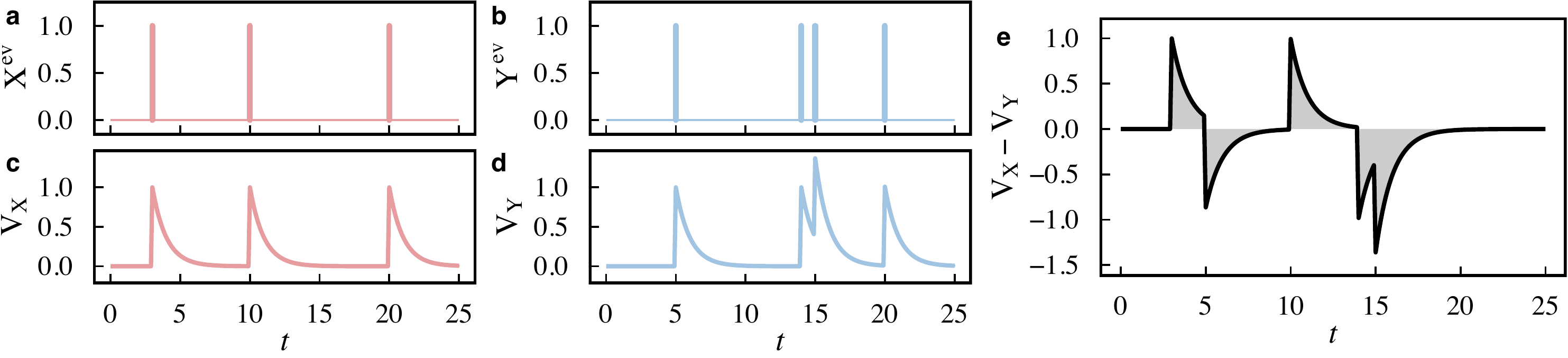}
\caption{\label{fig:van_rossum} General idea behind the van-Rossum distance \citep{rossum01}. Two event time series (a) $X^{ev}$ and (b) $Y^{ev}$ where vertical segments represent the events in both time series. The von-Rossum distance transforms each event into a real function illustrated respectively in (c) and (d). (e) Considering the difference between the two transformed time series $V_X - V_Y$, the von-Rossum distance corresponds to the area under this curve (light gray).}
\end{figure}

The van Rossum distance \citep{rossum01} is another method used to find synchronizations between events. The main idea is to transform each event into a real function and compare them, as illustrated in Fig.~\ref{fig:van_rossum}. The first step consists on mapping an event sequence $X' = \{t_1,\cdots,t_N\}$ into a function $V(X') = 1/N\sum_{i=1}^{N}h_{\tau}(t-x_i)u(t-x_i)$, where $h_{\tau}(t)$ is a filtering function. Examples of filtering functions include the Gaussian $h_{\tau}^g(t) = \exp(-t^2/(2\tau^2))/\sqrt{2\pi \tau^2}$ and the laplacian $h_{\tau}^l(t) = \exp(-|t|/\tau)/(2\tau)$ kernels. The parameter $\tau$ defines the co-occurrence time scale. The van Rossum distance can be calculated as

\begin{equation}
    \label{eq:vanrossum_dist}
    d_{\text{vr}}(X',Y') = \sqrt{\int_{0}^{\infty} [V(X') - V(X')]^2 \dif t}.
\end{equation}

One common approach to test the significance of event-based distance functions considers surrogate data \citep{boers19, ferreira21b}. The idea is to generate artificial event sequences by shuffling the events in the two original time series and then measuring their distance. After repeating this process a considerable high number of times, it is possible to construct a distribution of distances and use it as a confidence level for the comparison.

\subsection{Transforming a set of time series into a network}
\label{subsec:tss_to_nets}

A network or a graph \citep{bondy08,barabasi16}, $G(V,E)$ is composed by a set of $n$ nodes (also called vertices) $V$ and a set of $m$ links (also called edges) $E$ connecting two nodes. The most common representation for a network is the adjacency matrix $A$ whose entry $a_{ij} = 1$ if the nodes $i$ and $j$ are connected or $a_{ij} = 0$ otherwise. A weighted network can also be represented by $A$. In this case, each entry $a_{ij} = w_{ij} \in \mathbb{R}_{\neq 0}$ corresponds to the weight $w_{ij}$ between nodes $i$ and $j$, or $a_{ij}$ = 0 if there is no link.

\begin{figure}[t!]
\centering
\includegraphics[width=1\linewidth]{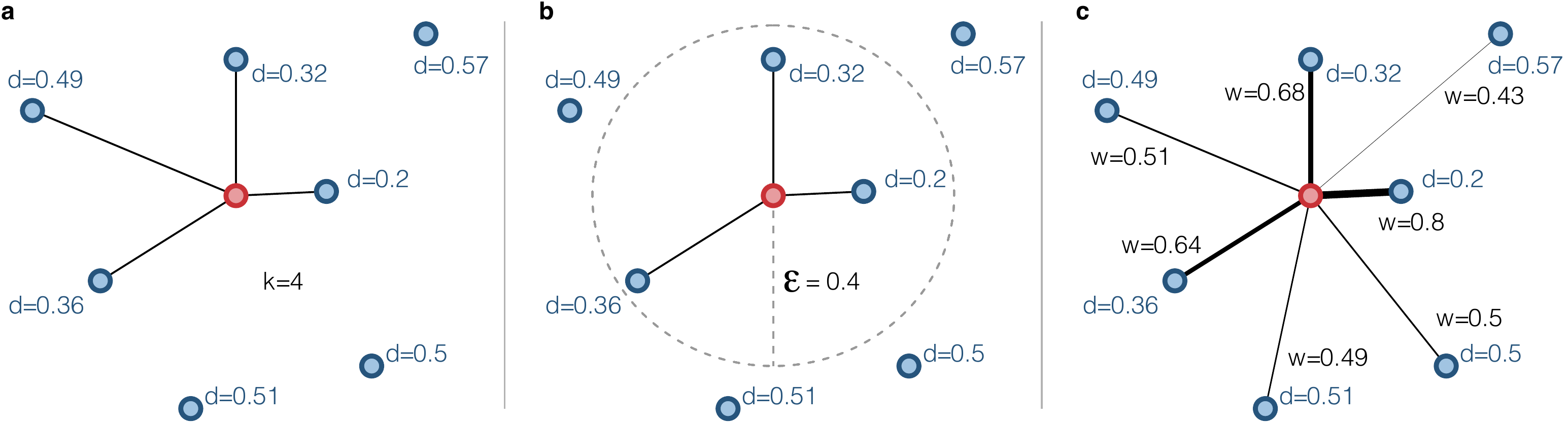}
\caption{\label{fig:nets_contruction} Geometrical intuition behind three different network construction processes. Each time series is represented by a blue node in the network, except the reference time series represented in red. The figures show the distances from the reference time series to all the other ones (a) In a $k$-NN ($k$=4) network, the reference node is connected to the four other ones with the shortest distances. (b) In a $\epsilon$-NN ($\epsilon$ = 0.4) network, the reference node is linked to all the nodes whose distance is shorter than $\epsilon$ (within the dashed circle). (c) In the weighted graph, the reference node is connected to all the other ones, and the link weight is ($w = 1 - d$). The link thickness is proportional to the weight.}
\end{figure}

A set of time series $\mathcal{X}$ can be transformed into a network by calculating the distance between all pairs of time series. These distances are represented by the distance matrix $D$ whose position $d_{ij}$ stores the distance between the time series $i$ and $j$. The network construction process consists of transform the distance matrix $D$ into an adjacency matrix $A$ where each node represents a time series. The most common building approaches are \citep{silva18,qiao18}:

\begin{itemize}
    \item \emph{$k$-NN network}: In a $k$ nearest neighbors network ($k$-NN), each node (representing a time series) is connected to the $k$ other ones with the shortest distances. Thus, the construction process requires finding for each row $i$ of $D$ the $k$ other elements with shortest distances, where $k \in \mathbb{W}$ is a parameter defined by the user. Approximation algorithms speed up the construction process  \citep{arya98,chen09}.
    
    \item \emph{$\epsilon$-NN network}: In a $\epsilon$ nearest neighbors network ($\epsilon$-NN), (Fig.~\ref{fig:nets_contruction}-b), each node is connected to all the other ones if their distances are shorter than $\epsilon \in \mathbb{R}_{+}$, defined by the user. An advantage of the $\epsilon$- over $\epsilon$-NN networks is that the $\epsilon$ permits a finer in the threshold choice. Another advantage is the fact that $\epsilon$ can restrict the connections to only strong ones, while the $k$-NN will always connect nodes regardless if the $k$ shortest distances are small or high. Similar to $k$-NN networks, approximations can speed up the construction process \citep{arya98,chen09}. 
    
    \item \emph{Weighted network}: Instead of limiting the number of links, a weighted network (Fig.~\ref{fig:nets_contruction}-c) can be constructed from $D$ by simply connecting all pairs of nodes and using their distances to set weights. In general, link weights have the opposite definition of a distance, which means that the shorter the distance between two time series, the stronger the link connecting the respective nodes. Considering that $D$ is in the interval [0,1], the weighted adjacency matrix can be defined as $A$ = 1 - $D$. Conversely, if $D \in [0,+\infty]$, $D$ can be be first normalized between [0,1] ($D_{\text{norm}}$) and then consider $A$ = 1 - $D_{\text{norm}}$.
    
    \item \emph{Networks with significant links}: In many applications, it is important to minimize the number of spurious links in the network. One way of minimizing these spurious links is by considering statistical tests for the distance functions. In other words, this network construction process connects all pairs of nodes only if their distance is statistically significant. For example, the significance of the Pearson correlation coefficient can be tested using the z-transformation (Eq.~\ref{eq:fisher_confidence_interval_z}) described in Sec.~\ref{seubsec:dist_funcs}. This network construction process can be seen as a special case of $\epsilon$-NN network where $\epsilon$ is a threshold for significant links. It is important to remember that this building approach minimizes the number of spurious links but does not guarantee that they will not appear. 
    
\end{itemize}

Networks can be static or temporal \citep{holme12,mikko14}. A static network $G(V,E)$ uses the whole historical data in a time series to construct a single network. Temporal networks divide the historical period into time windows that are used to construct layers in a temporal (multilayered) network. A temporal network is a sequence of static networks ordered in time. Each layer can be seen as a snapshot of the time window. The temporal network construction process requires the computation of a set of distance matrices $\mathcal{D} = \{D_1,\cdots,D_W\}$, one for each of the $W$ windows. A time window is defined by the user and has a length $\Delta t$ and a step $s$ that defines the number of values that are included and excluded in each window. The temporal network construction consists of transforming each distance matrix $D \in \mathcal{D}$ into a network layer using the same process used for a static network previously described in this section. One of the main advantages of this approach is to make possible the temporal analysis (evolution) of the modeled system. 


The network construction process has a combinatorial nature. Considering a set of $n$ time series, the whole construction process requires ${n\choose 2}$ distance calculations in a static network or $W\cdot{n\choose 2}$ in a temporal network, where $W$ is the number of layers. The transformation from the distance matrix to the adjacency matrix requires searches for pairs of shortest nodes, a task that is much faster than the distance calculations. For this reason, the conversion step ($D \to A$) can be neglected in the total time complexity is $O({n\choose 2})$. This means that the number of distance computations can increase very fast for very large time series data sets, making the network modeling approach infeasible.

\subsection{Transforming a single time series into a network}
\label{subsec:ts_to_nets}

This section presents methods to transform a single time series into a network \citep{zou19,silva21}. 

\subsubsection{Proximity networks}

The most straightforward method to perform this transformation consists of breaking the time series into time windows and using the same approach described for multiple time series (Sec.~\ref{subsec:tss_to_nets}). Each time window is considered a time series and the network construction process requires calculating a distance matrix $D$ between all time windows using a distance function. Then, $D$ can be transformed into an adjacency matrix using methods such as $k$-NN and $\epsilon$-NN networks \citep{silva18}. The time window size should be carefully chosen. For example, time windows too small can be problematic when calculating correlation distances. For this reason, this simple construction procedure may not be suitable for short time series because it could generate small time windows. 

\subsubsection{Transition networks}

\begin{figure}[t!]
\centering
\includegraphics[width=1\linewidth]{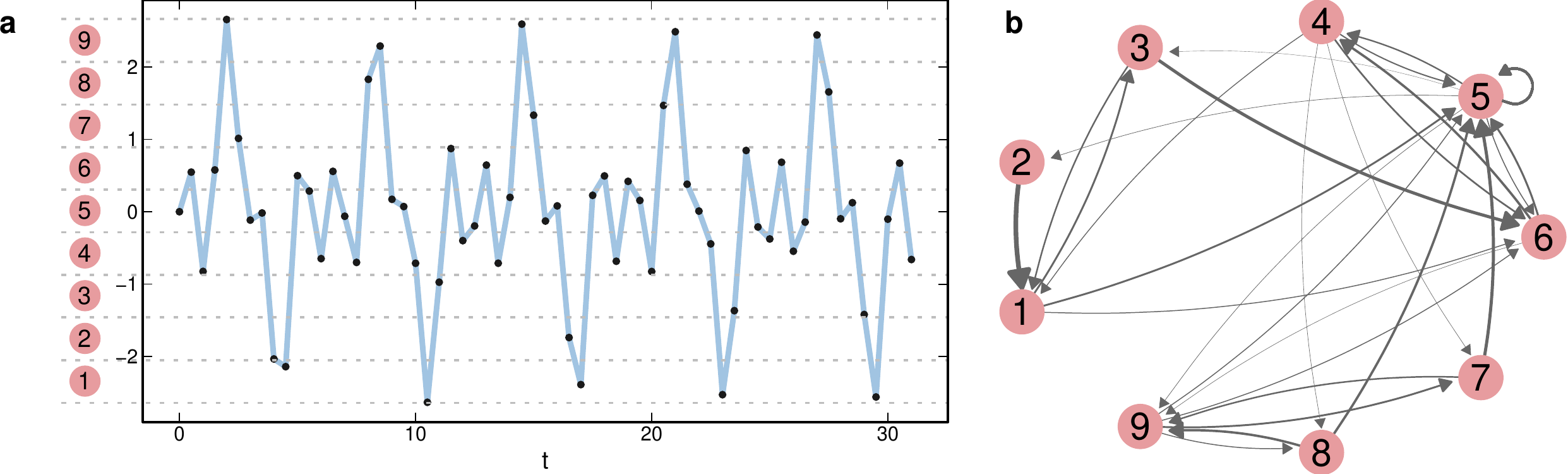}
\caption{\label{fig:transition_nets} Transition networks. (a) Example of time series sampled from $y(t) = \sin(t) + \sin(4t) - \sin(2t)$. This example uses nine equally-spaced bins divided by the dashed lines. Each bin is represented by a node depicted in red. (b) The resulting transition network is illustrated with edge widths proportional to the transition probability.}
\end{figure}

Transition networks map time series into symbols, represented by nodes and edges to link transitions between symbols. \citet{campanharo11} proposed a straightforward way of defining the symbols, which consists of first dividing a time series into equally-spaced bins. Each bin corresponds to a symbol that is mapped to a node in the network. Each time series value is mapped to a sequence of symbols according to the bins they are inserted. Then, for every pair of consecutive symbols, a directed link is established between the corresponding nodes. Two equal consecutive symbols lead to self-loops in the network. Edge weights count the number of times that a transition between the two symbols occurs. After the construction process, the weights of the outgoing links from each node can be normalized and interpreted as a transition probability between symbols. In this case, the network can be seen as a Markov Chain. Fig.~\ref{fig:transition_nets} presents an example of a transition network.

\subsubsection{Visibility graphs}

\begin{figure}[t!]
\centering
\includegraphics[width=1\linewidth]{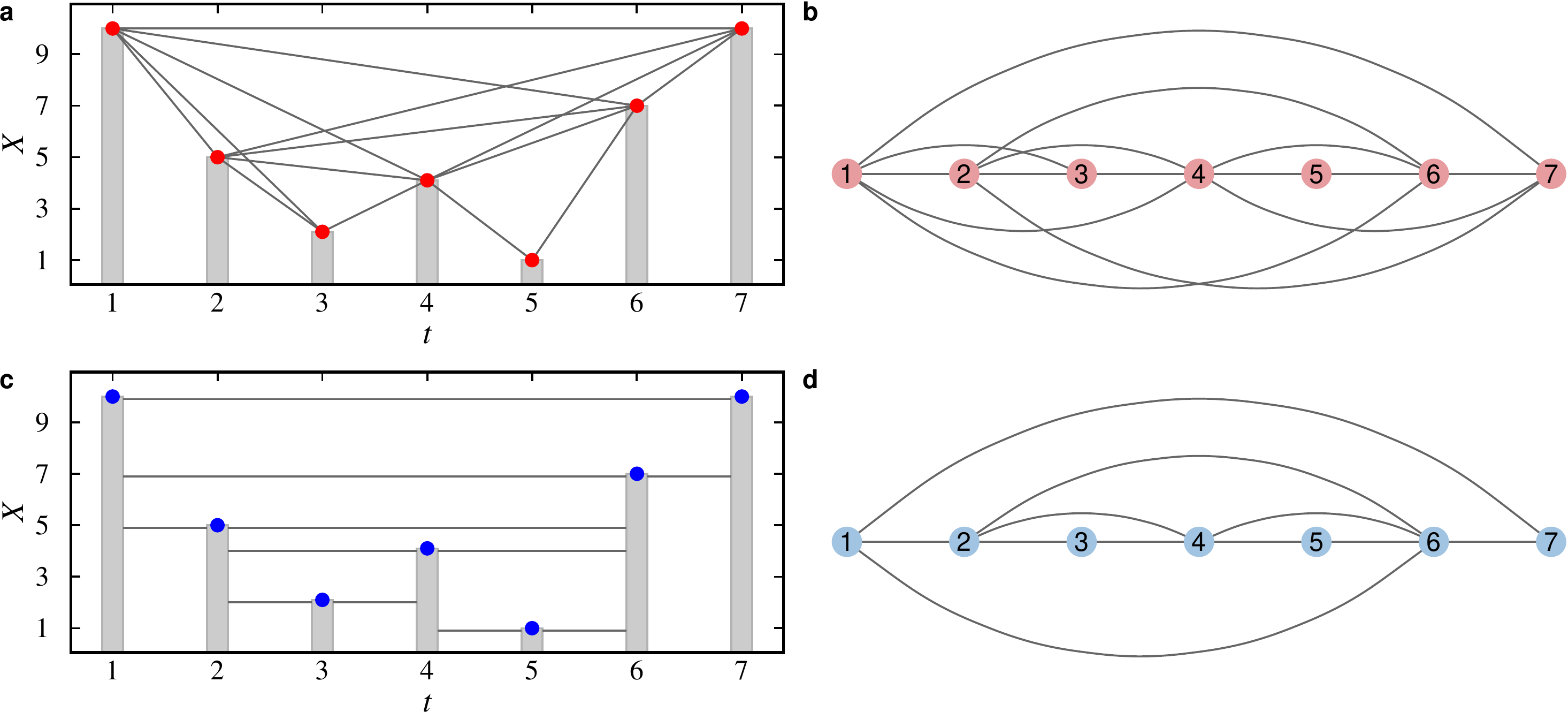}
\caption{\label{fig:visibility_graphs} Visibility graph construction. (a and c) An example of time series with values represented by the bars and points. Gray lines connect ``visible'' values as defined in the (a) natural (red) and (c) horizontal (blue) visibility graphs. The resulting natural (b) and horizontal (d) visibility graphs.}
\end{figure}

\citet{lacasa08} proposed a method called visibility graph that transforms a time series into a network. The idea is to represent each value in a time series by a bar in a barplot, as illustrated in Fig.~\ref{fig:visibility_graphs}. Then, each value becomes a node and links connect them if they can ``see'' each other. In the original definition, called natural visibility graph (NVG), the concept of visibility is defined as a line in any direction that does not cross any other bar. Considering a time series $X = \{X_i,\cdots,X_N\}$ with $N$ values and its natural visibility graph $NVG^{X}=\{v_i,\cdots,v_N\}$, two nodes $v_i$ and $v_j$ are connected if

\begin{equation}
    \label{eq:nvg}
    X_k < X_i + (X_j - X_i)\frac{k-i}{j-k}\ \forall \ k \in \{i < k < j\}.
\end{equation}

Other works adapted or restricted the idea of visibility. The horizontal visibility graph (HVG) \citep{luque09} defines visibility as horizontal lines between bars (Fig.~\ref{fig:visibility_graphs} c and d). In this case, two nodes $v_i$ and $v_j$ from $HVG^{X}$ are connected if

\begin{equation}
    \label{eq:hvg}
    X_k < X_i,X_j \ \forall \ k \in \{i < k < j\}.
\end{equation}

Threshold values can limit the maximum temporal distance between values \citep{ting12}. Visibility graphs can also be directed, representing the temporal order. Another important concern is the time complexity of the construction process. The original algorithms to construct NVGs and HVGs have quadratic time complexity O($n^2$), where $n$ is the number of elements in the series. Further improvements can reduce this time complexity to O($n\log n$) \citep{lan15}.

\subsubsection{Recurrence networks}

The transformation approach proposed by \citet{donner10} relies on the concept of recurrences in the phase space. The general idea consists of interpreting the recurrence matrix of a time series as an adjacency matrix. Nodes represent states in the phase-space and links connect recurrent ones. A recurrence matrix connects different points in time if the states are closely neighbored in phase space. Considering a time series $X$ and its $m$-dimensional time delay embedding $\vec{x} = (X_t, X_{t+\tau},\cdots,X_{t+(m-1)\tau})$ where $\tau$ is the delay, a recurrence matrix can be defined as

\begin{equation}
    \label{eq:recurrence_matrix}
    R_{ij}(\epsilon) = \Theta(\epsilon - d(\vec{x}_i - \vec{x}_j)),
\end{equation}

\noindent where $\Theta(\cdot)$ is the Heaviside function, $d(\cdot)$ is a distance function (e.g., Euclidean or Manhattan), and $\epsilon$ is a threshold distance that defines close states. A recurrence matrix $R(\epsilon)$ can then be transformed into an adjacency matrix $A$ of a network as:

\begin{equation}
    \label{eq:recurrence_net}
    A_{ij} = R_{ij}(\epsilon) - \delta_{i,j},
\end{equation}

where $\delta_{i,j}$ is the Kronecker delta introduced to remove self-loops. The resulting network is unweighted, undirected, and without self-loops 

\section{The ts2net Package}
\label{sec:ts2net}

The goal of the \pkg{ts2net} package is to provide methods to model a single or a set of time series into a network in \proglang{R}. It implements time series distance functions, parallel distance calculations, network construction methods, and some useful functions. Some functions are fully implemented, while others are adapted from other packages. Thus, another goal of the \pkg{ts2net} package is to put together useful methods from other packages.

The \pkg{ts2net} package is available on the Comprehensive \proglang{R} Archive Network (CRAN) and can be installed with the following command:
\begin{CodeChunk}
\begin{CodeInput}
R> install.packages("ts2net")
\end{CodeInput}
\end{CodeChunk}
Another option is to install \pkg{ts2net} from GitHub using the \fct{install\_github} function from either \pkg{remotes} or \pkg{devtools} packages:
\begin{CodeChunk}
\begin{CodeInput}
R> install.packages("remotes") # if not installed
R> remotes::install_github("lnferreira/ts2net")
\end{CodeInput}
\end{CodeChunk}
\subsection{Implementation aspects} 

The \pkg{ts2net} package relies on some other important packages. One of the most important ones is \pkg{igraph}, which is used to model networks \citep{package_igraph}. The \pkg{igraph} package provides many different graph and network analysis tools, which can be used to extract information from the modeled time series. The \pkg{igraph} objects can be easily transformed or exported to other formats such as graphml, gml and edge list. Besides, other network analysis packages commonly accept these objects. Examples are the \pkg{sna} \citep{package_sna}, \pkg{tsna} \citep{package_tsna}, and \pkg{muxViz} \citep{package_muxviz}.

The network construction process normally requires a high number of calculations. For this reason, the most important functions in \pkg{ts2net} are implemented in parallel using multi-core programming. The parallelization was implemented using the \fct{mclapply} function from the \pkg{parallel}, which comes with \proglang{R}. This implementation uses forks to create multiple processes, an operation that is not available in Windows systems. It means that, on Windows, all functions in \pkg{ts2net} with \texttt{num\_cores} can only run with one core.

Besides the multicore parallelization, the \pkg{ts2net} was also constructed to be run in supercomputers and computer clusters. The idea is to divide the distance calculations into jobs that could be calculated in parallel and then merged together to construct the distance matrix. The \fct{ts\_dist\_dir} and \fct{ts\_dist\_dir} functions make this process easier.

\subsection{Distance calculation}
\label{sec:ts2net_distance_calculation}

Functions to calculate the distance matrix:

\begin{itemize}
    \item \fct{ts\_dist}: Calculates all pairs of distances and returns a distance matrix $D$. This function receives a list of time series which can either have or not have the same length. It runs serial or in parallel (except in Windows) using \fct{mclapply} from \pkg{parallel} package. This function receives a distance function as a parameter. \pkg{Ts2net} provides some distance functions (functions starting with \fct{tsdist\_...}). Other distances can be easily implemented or adapted to be used within this function.
    
    \item \fct{ts\_dist\_part}: Calculate distances between pairs of time series (similarly to \fct{ts\_dist}) in part of a list. This function is particularly useful to run in parallel as jobs in supercomputers or computer clusters (HPC). Instead of having a single job calculating all distances (with \fct{ts\_dist}), this function permits the user to divide the calculation into different jobs. The result is a data frame with the distances $d_{ij}$ and indexes $i$ and $j$. To construct the distance matrix $D$, the user can merge all the results using the functions \fct{dist\_parts\_merge} or \fct{dist\_parts\_file\_merge}.
    
    \item \fct{ts\_dist\_part\_file}: This function works similar as \fct{ts\_dist\_part}. The difference is that it reads the time series from RDS files in a directory. The advantage of this approach is that it does not load all the time series in memory but reads them only when necessary. This means that this function requires much less memory and should be preferred when memory consumption is a concern, e.g., huge data set or very long time series. The disadvantage of this approach is that it requires a high number of file read operations which considerably takes more time during the calculations. The user can merge all the results using the functions \fct{dist\_parts\_merge} or \fct{dist\_parts\_file\_merge}.
    
\end{itemize}

List of distance functions:

\begin{itemize}
   
    \item \fct{tsdist\_cor}: Absolute (Eq.~\ref{eq:d_cor_abs}), positive (Eq.~\ref{eq:d_cor_pos}) or negative (Eq.~\ref{eq:d_cor_neg}) correlation distances. This function optionally test the significance for the correlation using the Fisher z-transformation (Eq.~\ref{eq:fisher_transform}) \citep{asuero06}.
    
   
    \item \fct{tsdist\_ccf}: Absolute (Eq.~\ref{eq:d_ccor_abs}), positive (Eq.~\ref{eq:d_ccor_pos}), or negative (Eq.~\ref{eq:d_ccor_neg}) cross-correlation distances.
   
    \item \fct{tsdist\_dtw}: Dynamic Time warping (DTW) distance (Sec.~\ref{subsubsec_dtw}). This function is a wrapper for the \fct{dtw} function from the \pkg{dtw} package \citep{giorgino09}.
    
    \item \fct{tsdist\_nmi}: Normalized mutual information distance (Eq.~\ref{eq:nmi_distance}). The implementation uses functions from the \pkg{infotheo} package \citep{package_infotheo}.
    
    \item \fct{tsdist\_voi}: Variation of information distance (Eq.~\ref{eq:voi}). The implementation uses functions from the \pkg{infotheo} package \citep{package_infotheo}.
    
    \item \fct{tsdist\_mic}: Maximal information coefficient (MIC) distance. This function transforms the \fct{mine} function from the \pkg{minerva} package \citep{package_minerva} into a distance.
    
    \item \fct{tsdist\_es}: The events synchronization distance (Eq.~\ref{eq:quiroga_count3}) proposed by \citet{quiroga02}. This function can also use the modification proposed by \citet{boers19} ($\tau_{\text{max}}$). This function optionally test the significance for the events synchronization by shuffling the events in the time series. 
    
    \item \fct{tsdist\_vr}: Van Rossum distance. This function uses the \fct{fmetric} function from the \pkg{mmpp} package \citep{hino15}. It optionally test the significance for the events synchronization by shuffling the events in the time series. 
    
    
\end{itemize}

Other distances can be easily applied in the \pkg{ts2net}. The only two requirements are that the distance function should accept two arrays and it must return a distance. The packages \pkg{TSclust} \citep{package_tsclust} and \pkg{TSdist} \citep{package_tsdist} provide many other distance functions that can be used to create network with \pkg{ts2net}. For example, the euclidean distance (\fct{diss.EUCL} from \pkg{TSclust}) can be used to construct the distance matrix:
\begin{CodeChunk}
\begin{CodeInput}
R> install.packages("TSclust")
R> library(TSclust)
R> D <- ts_dist(ts_list, dist_func=diss.EUCL)
\end{CodeInput}
\end{CodeChunk}
It is important to remember that the parameter \code{dist_func} receives a function, so no parenthesis is necessary (\code{dist_func=diss.EUCL()} is an error). 

Another possibility is to implement specific functions and apply them. For example, the following code snippet implements a distance function that compares the mean values of each time series and applies it to calculate the distance matrix $D$:
\begin{CodeChunk}
\begin{CodeInput}
R> ts_dist_mean <- function(ts1, ts2) abs(mean(ts1) - mean(ts2))
R> D <- ts_dist(ts_list, dist_func=ts_dist_mean)
\end{CodeInput}
\end{CodeChunk}

\subsection{Network construction methods}

Multiple time series into a network:

\begin{itemize}
    
    \item \fct{net\_knn}: Constructs a $k$-NN network from a distance matrix $D$.
    
    \item \fct{net\_knn\_approx}: Creates an approximated $k$-NN network from a distance matrix $D$. This implementation may omit some actual nearest neighbors, but it is considerably faster than \fct{net\_knn}. It uses the \fct{kNN} function from the \pkg{dbscan} package.
    
    \item \fct{net\_enn}: Constructs a $\epsilon$-NN network from a distance matrix $D$.
    
    \item \fct{net\_enn\_approx}: Creates an approximated $\epsilon$-NN network from a distance matrix $D$. This implementation may omit some actual nearest neighbors, but it is considerably faster than \fct{net\_enn}. It uses the \fct{frNN} function from the \pkg{dbscan} package.
    
    \item \fct{net\_weighted}: Creates a full weighted network from a distance matrix $D$. In this case, $D$ should be in the interval $[0,1]$ (see \fct{dist\_matrix\_normalize}). The link weights in the resulting network is $w_{ij} = 1 - d_{ij}$. 
    
    \item \fct{net\_significant\_links}: Construct a network with significant links. 
    
\end{itemize}

A single time series into a network:

\begin{itemize}

    \item \fct{ts\_to\_windows}: Extract time windows from a time series. This function can be used to construct a network from time windows. It returns a list of windows that can be used as input for \fct{ts\_dist} to construct a distance matrix $D$.
    
    \item \fct{tsnet\_qn}: Creates transition (quantile) networks. 
    
    \item \fct{tsnet\_vg}: Creates natural and horizontal visibility graphs. This function can be run in parallel. 
    
    \item \fct{tsnet\_rn}: Creates a recurrence network from a time series. This function uses the \fct{rqa} function from the \pkg{nonlinearTseries} package to obtain the recurrence matrix. 
    
\end{itemize}

\subsection{Useful functions}

\begin{itemize}
    \item \fct{dist\_matrix\_normalize}: normalizes (min-max) a distance matrix $D$ between zero and one.
    
    \item \fct{dist\_percentile}: Returns the distance value that corresponds to the desired percentile. This function is useful when the user wants to generate networks with different distance functions but with the same link density.
    
    \item \fct{ts\_dist\_dirs\_merge}: \fct{ts\_dist\_dir} and \fct{ts\_dist\_dir} calculate parts of the distance matrix $D$. This function merges the results and constructs a distance matrix $D$.
    
    \item \fct{tsdist\_file\_parts\_merge}: Merges the distance parts as \fct{ts\_dist\_dirs\_merge}, but reading them from files. 
    
    \item \fct{events\_from\_ts}: Function to extract events from a time series. For example, it can return the desired percentile of highest or lowest values. This function is particularly useful for event-based distance functions.
    
\end{itemize}

\subsection{Data}

\begin{itemize}

    \item \fct{dataset\_sincos\_generate}: Generates a data set with sine and cosine time series with arbitrary noise.

    \item \fct{random\_ets}: Creates random event time series with uniform probability.

    \item \texttt{us\_cities\_temperature\_df}: A data set containing the temperature in 27 US cities between 2012 and 2017. This data set was adapted from the original \citep{kaggle_dataset} for simplicity reasons. Different from the original, this data set comprehends only cities in the US, grouped by month (mean value) temperature and without days with missing data. The data set is stored in a data frame with 61 rows and 28 variables where each column (except date), corresponds to the mean temperature values (\textdegree{}C) for each city.

    \item \texttt{us\_cities\_temperature\_list}: Similar to \texttt{us\_cities\_temperature\_df}, but in a list, which is the required format for distance calculations (e.g. \fct{ts\_dist}).

\end{itemize}

\subsection{Source code}

The \pkg{ts2net} is open-source and free to use according to the MIT License (\url{https://github.com/lnferreira/ts2net/blob/main/LICENSE}). All the code is available on Github at
\url{https://github.com/lnferreira/ts2net}.


\section{Applications} 
\label{sec:applications}

This section presents how the \pkg{ts2net} package can be used to transform a set or a single time series into networks. Two real data sets were used as examples. The goal here is not to present complicated applications and deep analyses but straightforward and didactic examples of how to use networks to model time series. 

\subsection{A set of time series into a network}

\begin{figure}[h!]
\centering
\includegraphics[width=1\linewidth]{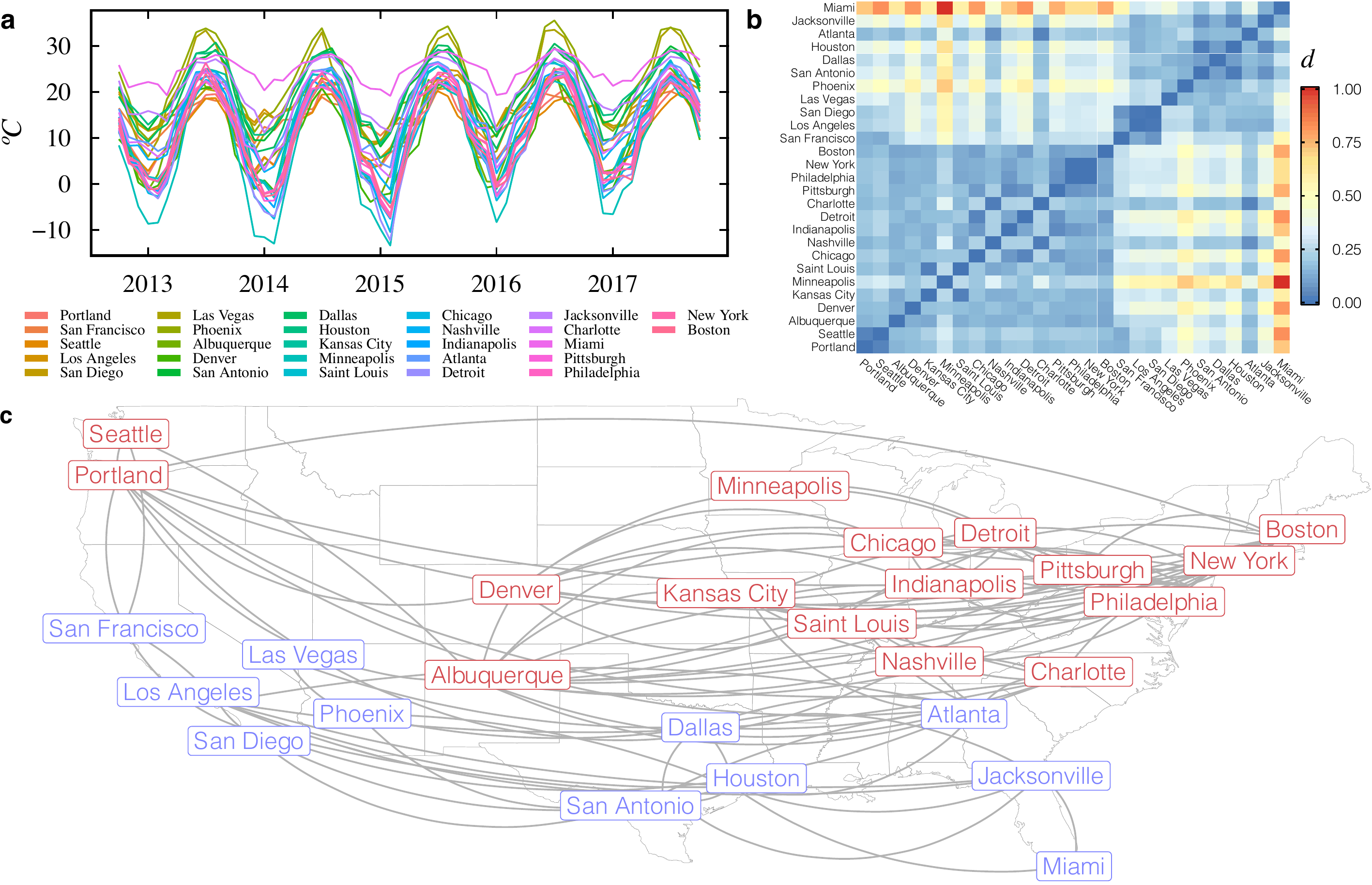}
\caption{\label{fig:ts2net_temperature} Transforming time series into a network using \pkg{ts2net}. (a) The historical temperature of 27 cities in the US. (b) The distance matrix $D$ (normalized DTW) for the data set. (c) The $\epsilon$-NN network was constructed using 30\% of the shortest distances. Node colors represent communities.}
\end{figure}

This section presents a network modeling example using real data. The data set is composed of monthly temperature data from 27 cities in the US between 2012 and 2017, illustrated in Fig.~\ref{fig:ts2net_temperature}-a. The version presented here was extracted and adapted from Kaggle \citep{kaggle_dataset} for simplicity reasons by taking the monthly averages and keeping only US cities. 

The following code calculates the distance matrix (Fig.~\ref{fig:ts2net_temperature}-b) using DTW, finds the $\epsilon$ that corresponds to the 30\% of the shortest distances, constructs a $\epsilon$-NN network (Fig.~\ref{fig:ts2net_temperature}-c), and find communities using the algorithm proposed by \citet{girvan02}:

\begin{CodeChunk}
\begin{CodeInput}
R> D <- ts_dist(us_cities_temperature_list, dist_func = tsdist_dtw)
R> eps <- dist_percentile(D, percentile = 0.3)
R> net <- net_enn(D, eps)
R> cluster_edge_betweenness(net)
\end{CodeInput}
\begin{CodeOutput}
IGRAPH clustering edge betweenness, groups: 2, mod: 0.33
+ groups:
  $`1`
   [1] "Portland"     "Seattle"      "Albuquerque"  "Denver"      
   [5] "Kansas City"  "Minneapolis"  "Saint Louis"  "Chicago"     
   [9] "Nashville"    "Indianapolis" "Detroit"      "Charlotte"   
  [13] "Pittsburgh"   "Philadelphia" "New York"     "Boston"      
  
  $`2`
   [1] "San Francisco" "Los Angeles"   "San Diego"    
   [4] "Las Vegas"     "Phoenix"       "San Antonio"  
   [7] "Dallas"        "Houston"       "Atlanta"      
  + ... omitted several groups/vertices
\end{CodeOutput}
\end{CodeChunk}

Fig.~\ref{fig:ts2net_temperature}-c illustrates the resulting network where nodes are conveniently placed according to their geographical position. Node colors represent the two communities that cluster together colder (blue nodes) and hotter (red nodes) cities.

\subsection{Transforming a set of time series into a network}
\label{subsec:application_tss_net}

\begin{figure}[h!]
\centering
\includegraphics[width=1\linewidth]{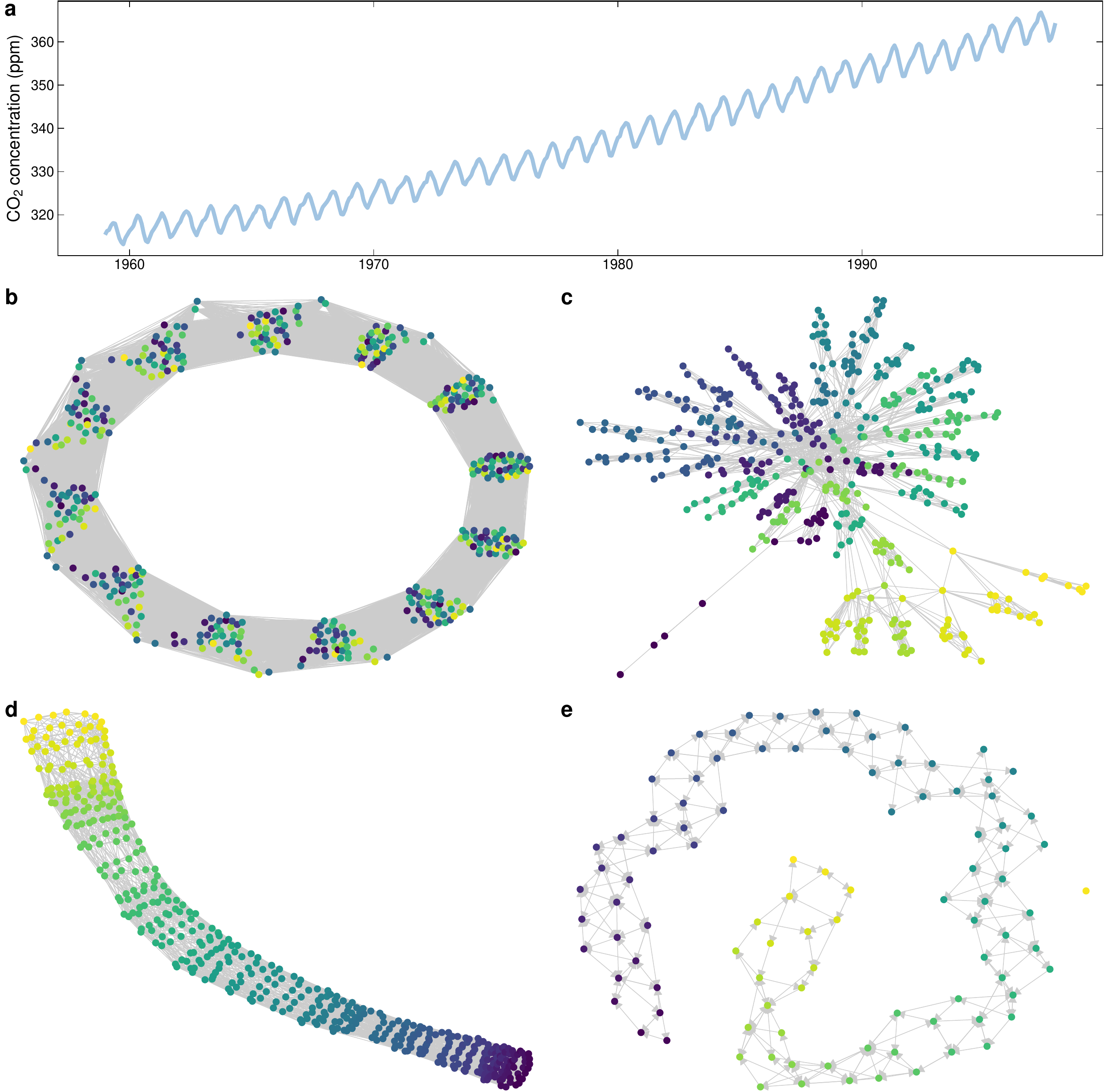}
\caption{\label{fig:ts2net_co2} Transforming a single time series into networks using \pkg{ts2net}. (a) The CO$_{2}$ concentration time series. (b) Proximity network with time window 12 and one-value step. (c) Natural visibility graph. (d) Recurrence network ($\epsilon = 5$). (e) Transition (quantile) network (100 equally-spaced bins). Node colors represent temporal order (yellow to blue), except in the transition network where colors represent the sequence of lower (yellow) to higher (blue) bins.}
\end{figure}

This section presents how a time series extracted from real data can be modeled as networks. The time series with the monthly atmospheric concentration of carbon dioxide (ppm) measured in the Mauna Loa Observatory between 1959 and 1997 \citep{keeling05}, illustrated in Fig.~\ref{fig:ts2net_co2}a. This data set was mainly chosen by its simplicity and accessibility (available with the base installation of R).

The following code generates a time-window network (\texttt{net\_w}) with width 12 and one-value step, a visibility graph (\texttt{net\_vg}), a recurrence network (\texttt{net\_rn}) with $\epsilon = 5$, and a transition network (\texttt{net\_qn}) using 100 equally spaced bins using the \pkg{ts2net} package. 

\begin{CodeChunk}
\begin{CodeInput}
R> net_w = as.vector(co2) |> 
       ts_to_windows(width = 12, by = 1) |> 
       ts_dist(cor_type = "+") |> 
       net_enn(eps = 0.25)
R> net_vg = as.vector(co2) |> 
       tsnet_vg()
R> net_rn = as.vector(co2) |>
       tsnet_rn(radius = 5)
R> net_qn = as.vector(co2) |> 
       tsnet_qn(breaks = 100)       
\end{CodeInput}
\end{CodeChunk}

Figures~\ref{fig:ts2net_co2}b-e illustrate the networks resulting from the transformation of the CO$_{2}$ time series. From the topology of the resulting networks, it is possible to extract useful information. For example, the time-window network (Fig.~\ref{fig:ts2net_co2}b) shows 12 groups representing each month. The visibility graph (Fig.~\ref{fig:ts2net_co2}c) presents many small-degree nodes that represent the values on valleys while a few hubs connect peaks values of CO$_{2}$ after 1990 that connect to many other nodes. These hubs appear due to the increasing trend of CO$_{2}$ contraction and some slightly higher peaks during the seasons after 1980. The recurrence (Fig.~\ref{fig:ts2net_co2}d) and transition (Fig.~\ref{fig:ts2net_co2}e) networks present some local connections caused by the seasons, and a line shape as a result the increasing trend in the CO$_{2}$ concentration over the years.

\section{Final considerations} \label{sec:summary}

This paper presents \pkg{ts2net}, a package to transform time series into networks. One important drawback when transforming a set of time series into a network is the high computational cost required to construct a distance matrix (combinations of time series), making it unfeasible for huge data sets. To minimize this problem, \pkg{ts2net} provides tools to run distance functions in parallel and in high-performance computers via multiple jobs. 

The \pkg{ts2net}'s goal is not to provide all transformation methods in the literature but the most used ones. It does not mean that this package cannot be extended or customized. For example, as explained in Sec.~\ref{sec:ts2net_distance_calculation}, this package accepts other time series distance functions implemented by other packages or the user. Additional modeling approaches can also be incorporated to the package in the future.


\section*{Computational details}


The \pkg{ts2net} package requires \proglang{R} version 4.1.0 or higher. \proglang{R} itself and all packages used are available from the Comprehensive \proglang{R} Archive Network (CRAN) at \url{https://CRAN.R-project.org/}.




\bibliography{refs}

\end{document}